\documentclass[journal]{IEEEtran}

\usepackage{amsmath}
\usepackage{amssymb}
\usepackage{amsthm}
\usepackage{graphicx}
\usepackage{booktabs}
\usepackage[table]{xcolor}
\usepackage{placeins}
\usepackage[hidelinks]{hyperref}
\usepackage{etoolbox}

\patchcmd{\abstract}{---}{\enspace}{}{\PackageError{rmsampler}{Abstract separator patch failed}{}}
\patchcmd{\IEEEkeywords}{---}{\enspace}{}{\PackageError{rmsampler}{Keyword separator patch failed}{}}

\newtheoremstyle{radiodiffroman}
  {3pt}
  {3pt}
  {\normalfont}
  {}
  {\normalfont}
  {.}
  {0.5em}
  {}
\theoremstyle{radiodiffroman}
\newtheorem{theorem}{Theorem}
\newtheorem{corollary}[theorem]{Corollary}
\newtheorem{proposition}[theorem]{Proposition}
\newenvironment{radioproof}
  {\par\pushQED{\qed}\normalfont\noindent Proof.\ }
  {\popQED\par}

\title{RMSampler: Sparse Radio Map Reconstruction with Active Sampling and Diffusion Posterior Correction}

\author{Nan Cheng and Xiucheng Wang%
\thanks{Nan Cheng and Xiucheng Wang are with the School of Telecommunications Engineering, Xidian University, Xi'an, China.}}
\hypersetup{pdfauthor={Nan Cheng and Xiucheng Wang}}

\begin{document}

\maketitle

\begin{abstract}
Reconstructing a radio map (RM) from sparse received-power measurements supports non-cooperative interference monitoring and spectrum management.
However, sparse measurements leave uncertainty about both the global RM and the interfering emitter location, making informative sampling and accurate reconstruction challenging.
To address these challenges, a novel joint sampling and reconstruction method named RMSampler is proposed in this paper to improve measurement selection and RM reconstruction under limited measurement budgets.
For measurement selection, RMSampler evaluates global RM error reduction using the covariance of diffusion-generated RMs and localization error reduction using candidate emitter locations.
Because these errors have different units, their expected relative reductions are combined into a scale-invariant sampling criterion, with each acquired measurement updating both uncertainty estimates.
After sampling, the updated emitter-location probabilities guide a separate conditional diffusion ensemble for final reconstruction.
To improve measurement agreement, an effective sample size (ESS) target guides robust posterior correction, followed by a minimum-distortion projection enforcing measurement and building constraints.
For the reconstruction ensemble, analysis establishes that increasing likelihood power reduces expected robust measurement loss, while pretrained neural parameters and architecture remain unchanged.
Experiments using simulated and real-world measured data demonstrate improved reconstruction under matched measurement budgets.
In the Chongqing real-world measured data experiment, root mean square error (RMSE) decreases from 3.54 to 2.73~dB relative to random sampling under a common reconstruction method.
\end{abstract}


\section{Introduction}
\label{sec:introduction}

Reconstructing a radio map (RM) from sparse received-power measurements enables non-cooperative interference monitoring across a region~\cite{romero2022radiomap}. By estimating received power at unmeasured locations, the RM identifies interference-affected areas and supports spectrum management~\cite{wang2026tutorial}. In non-cooperative scenarios, the interfering emitter location and propagation parameters can be unavailable, making receiver measurements and environmental information central to reconstruction~\cite{11466421}. Acquiring these measurements consumes sensing time and equipment resources, so exhaustive measurement campaigns are costly~\cite{polyzos2024bayesian}. A limited measurement budget therefore requires informative sampling locations and an inference method that effectively uses the acquired values.

However, sparse measurements leave uncertainty in both the global RM and the interfering emitter location, creating a joint estimation problem for measurement selection. Building blockage and multipath propagation produce received-power variations that simple distance-based models cannot capture~\cite{levie2021radiounet}. Learned propagation priors help reconstruct these variations from sparse observations~\cite{teganya2022deepcompletion}, and diffusion-based inverse estimation reconstructs the RM when the emitter location is unknown~\cite{11466421}. For selecting further measurements, predictive uncertainty guides Gaussian process (GP) active learning~\cite{polyzos2024bayesian} and flow-matching-based RM sensing~\cite{sun2026flowactive}. Evaluating the resulting reconstruction benefit across the RM additionally requires the spatial dependence between a candidate measurement and unmeasured locations~\cite{krause2008sensor}. Mobile GP sensing also addresses RM reconstruction and emitter localization through confidence-based acquisition~\cite{zhang2025adaptive}. These estimation objectives motivate a common criterion that evaluates how the same measurement reduces global RM error and emitter-location error.

Normalizing the expected reductions in these errors allows both estimation objectives to guide the next sampling decision on a common scale. Candidate RMs from a pretrained diffusion model~\cite{wang2025radiodiff} provide spatial covariance between a candidate measurement and unmeasured RM locations. Updating the candidate weights with acquired measurements makes this covariance reflect the current reconstruction uncertainty. Under a Gaussian approximation, it predicts the candidate measurement's reduction in global RM error. For emitter localization, received-power predictions under candidate emitter locations quantify how the same observation could reduce expected coordinate error. Dividing each predicted reduction by the corresponding current expected error accounts for their different units and magnitudes. The normalized sum therefore ranks sampling locations by their joint estimation benefit, and each acquired value updates the RM weights and emitter-location probabilities used for the next decision.

\begin{figure}[!t]
\centering
\includegraphics[width=\columnwidth]{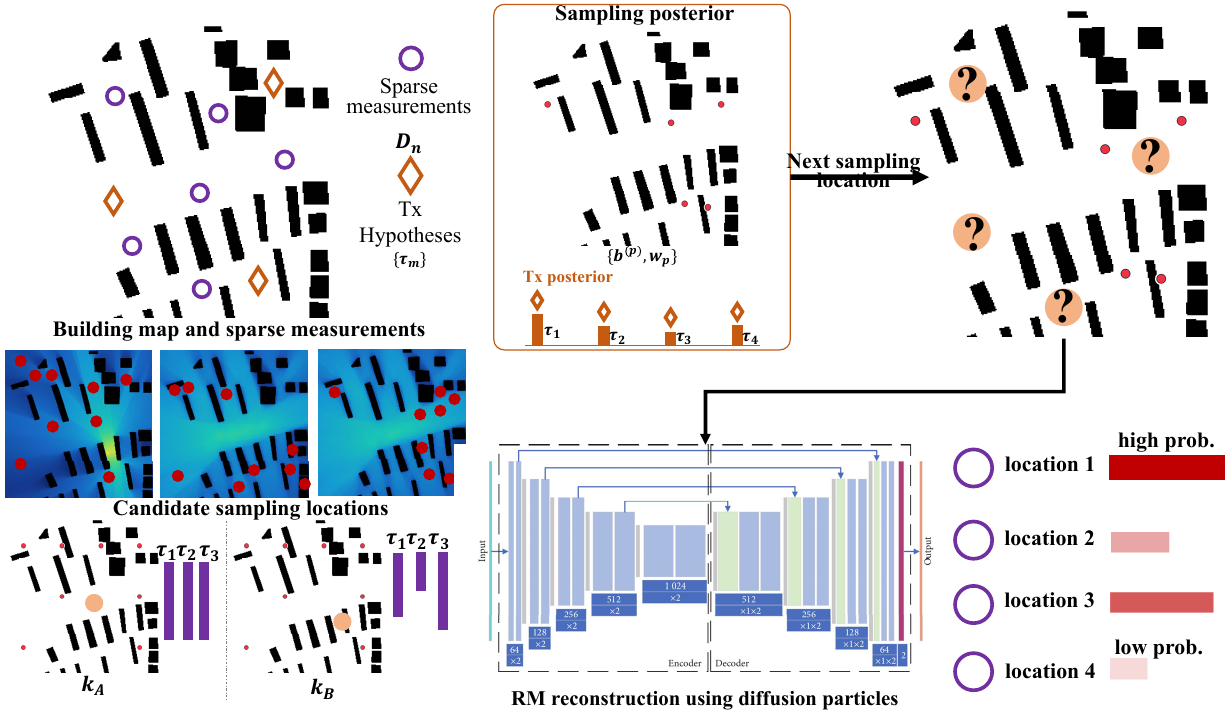}
\caption{RMSampler sampling and reconstruction using the building map, received-power measurements, and transmitter (Tx) location hypotheses.}
\label{fig:system_overview}
\end{figure}

To implement this joint criterion, a novel sampling and RM reconstruction method named RMSampler is proposed in this paper to improve measurement selection and reconstruction under a limited measurement budget. The criterion guides sequential sampling while the acquired values update the probabilities of candidate emitter locations. These probabilities then condition a separate diffusion ensemble for final RM reconstruction, connecting the information obtained during sampling to the reconstruction prior. To improve agreement with the acquired measurements, the generated RMs are reweighted by a robust measurement likelihood. An effective sample size (ESS) target guides correction strength to balance measurement agreement and weight concentration, after which a minimum-distortion projection enforces bounded measurement and building constraints. This measurement-dependent inference uses both pretrained models without changing their neural parameters or architectures, as illustrated in Fig.~\ref{fig:system_overview}. The main contributions of this paper are summarized as follows.

\begin{enumerate}
\item We formulate measurement-budgeted RM reconstruction as a joint sequential estimation and sampling problem with an unknown interfering emitter location. Each acquired measurement updates RM reconstruction uncertainty and emitter-location uncertainty, allowing both estimation objectives to inform subsequent sampling and final reconstruction.
\item We derive a sampling criterion from fractional reductions in expected global RM error and emitter-location coordinate error. Spatial covariance evaluates the reconstruction contribution, while candidate emitter locations evaluate the localization contribution. We establish invariance to changes in received-power scale and offset when the corresponding variances are rescaled. Under Gaussian and finite-hypothesis approximations, maximizing the criterion minimizes a clipped normalized one-step joint estimation error.
\item We develop ESS-guided robust posterior correction for a separately generated conditional RM ensemble. For a fixed ensemble, we derive a Kullback-Leibler (KL)-regularized variational formulation and establish that increasing likelihood power reduces expected robust measurement loss. An ESS-guided search selects correction strength, and a minimum-distortion projection enforces bounded measurement and building constraints while preserving both pretrained diffusion models.
\item We evaluate RMSampler on 100 RadioMapSeer test scenarios~\cite{levie2021radiounet}, Vienna real-world measured data~\cite{wiedner2026vienna}, and Chongqing real-world measured data~\cite{xu2026complex}. Simulated RM comparisons evaluate normalized mean squared error (NMSE), structural similarity index (SSIM), and emitter localization. RMSampler, Uniform, and Random share a conditional diffusion reconstructor to isolate sampling gains, while measured-data controls separately assess reconstruction and sampling. In the Chongqing real-world measured data experiment on New Radio cell identity (NCI) groups, root mean square error (RMSE) decreases from 3.54 to 2.73~dB relative to random sampling using the same reconstruction method and a budget of 16 measurements.
\end{enumerate}

\begin{figure}[!t]
\centering
\includegraphics[width=\columnwidth]{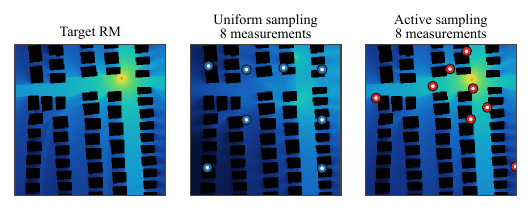}
\caption{Selected test example with eight measurements and a common conditional diffusion reconstructor.}
\label{fig:motivation_locations}
\end{figure}

\section{Related Work}
\label{sec:related_work}

RM reconstruction combines sparse measurements with spatial dependencies learned or specified before inference. Interpolation and graph regularization exploit these dependencies directly~\cite{romero2022radiomap}, while neural completion learns propagation patterns from other environments~\cite{teganya2022deepcompletion}. Low-rank neural models extend completion to multidimensional radio data~\cite{shrestha2022deepspectrum}. For urban RMs, RadioUNet learns the effect of geometry on received power~\cite{levie2021radiounet}, and DeepREM estimates unmeasured values from sparse observations~\cite{chaves2023deeprem}. Conditional adversarial learning provides another data-driven reconstruction approach~\cite{zhang2023rmegan}. These methods supply reconstruction mechanisms, while active sampling additionally requires the expected benefit of a measurement before its value is observed.

Generative priors represent multiple reconstructions consistent with incomplete observations. Learned generative representations constrain inverse recovery~\cite{bora2017compressed,mardani2019deepgan}, and diffusion and score-based models generate samples through iterative denoising~\cite{ho2020ddpm,song2021score}. Latent diffusion performs this generation in a learned compressed representation~\cite{rombach2022latent}. In RM construction, RadioDiff supplies a geometry-conditioned propagation prior~\cite{wang2025radiodiff}. Its extensions address three-dimensional RMs~\cite{11083758}, multipath structure through the Helmholtz equation~\cite{11278649}, and Bayesian recovery from noisy sparse measurements~\cite{11466421}. The broader relationship between forward prediction, inverse reconstruction, and physical information is reviewed in~\cite{wang2026tutorial}. Diffusion inverse solvers incorporate measurement consistency through manifold corrections~\cite{chung2022manifold}, denoising restoration~\cite{kawar2022ddrm}, likelihood-guided posterior sampling~\cite{chung2023dps}, or pseudoinverse guidance~\cite{song2023pseudoinverse}. These approaches use observations at specified locations, motivating an additional decision rule for selecting the next measurement.

Active sampling assesses a measurement before acquisition through a specified estimation objective. Bayesian experimental design formalizes this decision~\cite{chaloner1995bayesian}, with expected-error reduction~\cite{cohn1996active}, mutual-information sensor placement~\cite{krause2008sensor}, and upper-confidence acquisition~\cite{srinivas2010gpucb} providing different criteria. Adaptive source localization uses observations to reduce location uncertainty~\cite{vergassola2007infotaxis}. For RM reconstruction, GP ensembles supply predictive uncertainty~\cite{polyzos2024bayesian}, and GP upper-confidence-bound sampling with cross-entropy planning (GP-UCB-CE) incorporates receiver motion~\cite{zhang2025adaptive}. Flow matching also uses generative diversity to guide measurement selection~\cite{sun2026flowactive}. Geometry-aware Cholesky projection (GACP) uses environmental geometry, and GACP with autoencoder reconstruction (GACP-AE) combines that selection with learned RM completion~\cite{chai2026geometry}. Energy-modified leverage sampling (EMLS) exploits low-rank structure~\cite{sun2024energy}. RMSampler evaluates the expected RM and emitter-coordinate error reductions of the same candidate observation, with normalization making their contributions comparable.

Real-world measured data allow these reconstruction priors to be evaluated under propagation conditions that differ from training. The Vienna dataset combines city-scale fourth-generation (4G) and fifth-generation (5G) radio measurements with high-resolution building and terrain models \cite{wiedner2026vienna}, and the Chongqing macro-cell dataset provides ground and aerial measurements with complex-terrain descriptors \cite{xu2026complex}. Their recorded receiver measurements and environmental data complement complete simulated RMs by supporting evaluation at actual receiver locations.

\section{System Model and Problem Formulation}
\label{sec:system_model}

Sequential sampling and RM reconstruction are linked through the posterior updated by each acquired measurement. The current posterior predicts a candidate measurement's expected benefit, while its acquired value updates the information used for subsequent sampling and reconstruction. We express this dependence through expected RM reconstruction error and expected emitter-location error. The building map is known throughout inference, and noisy outdoor received-power measurements arrive sequentially without access to the true emitter coordinate.

\subsection{Scenario and Measurement Model}

We consider a static single-emitter scenario on a grid with height $H$, width $W$, and $D=HW$ pixels. The binary building map $\boldsymbol{m}\in\{0,1\}^{D}$ is accurate and available throughout inference, where $m_i=1$ identifies a building pixel. The building set is $\mathcal{B}=\{i:m_i=1\}$, and the set of admissible outdoor locations is $\mathcal{O}=\{1,\ldots,D\}\setminus\mathcal{B}$. Building-pixel values $v_i$ follow the known map convention and are provided as side information without consuming the sampling budget. The unknown emitter coordinate $\tau$ belongs to $\mathcal{O}$, and the RM $x\in[0,1]^D$ contains normalized received power. On the outdoor region, we decompose the RM into a background component $b$ and an emitter-dependent component $s(\tau)$, which gives
\begin{align}
    x_i & =
    \begin{cases}
        b_i+s_i(\tau), & i\in\mathcal{O},\\
        v_i, & i\in\mathcal{B}.
    \end{cases}
    \label{eq:rm_decomposition}
\end{align}
The decomposition in \eqref{eq:rm_decomposition} separates background received-power variation from variation associated with the emitter location. Diffusion particles approximate the former, while a calibrated propagation model predicts the received-power changes associated with alternative emitter coordinates. Each new measurement can therefore update both candidate RM weights and emitter-location probabilities through their respective predictions.

A mobile sensor sequentially acquires noisy received-power measurements at distinct outdoor pixels. Let $k_n\in\mathcal{O}$ be the location selected at round $n$, and let $y_n$ denote the corresponding measurement. The accumulated data after round $n$ are $\mathcal{D}_n=\{(k_j,y_j)\}_{j=1}^{n}$. Sensor noise has known variance $\sigma_{k_n}^{2}$ and satisfies $\mathbb{E}[\epsilon_n]=0$ and $\mathbb{E}[\epsilon_n^2]=\sigma_{k_n}^{2}$. A robust likelihood accounts separately for model discrepancy, giving the measurement model
\begin{align}
    y_n
    &=x_{k_n}+\epsilon_n
      =b_{k_n}+s_{k_n}(\tau)+\epsilon_n.
    \label{eq:measurement_model}
\end{align}
The sampling budget counts one received-power measurement per selected outdoor location. All candidate locations have the same measurement cost, so the design selects a sequence of distinct sampling locations under this budget. Removing each selected pixel from the candidate set prevents repeated sampling. Both sampling and reconstruction use only the acquired values and available side information, with the true emitter coordinate and unobserved outdoor values excluded.

\subsection{Joint Posterior and Decisions}

The building map and sparse measurements induce uncertainty over both the RM and the emitter coordinate. Let $q_{\theta}(x\mid \boldsymbol{m},r(\tau))$ denote the frozen RadioDiff prior conditioned on the building map and a rasterized emitter-location hypothesis $r(\tau)$. Let $\pi_0(\tau\mid \boldsymbol{m},v)$ denote the initial finite spatial distribution produced from the building map available at inference. RMSampler fuses these objects through a modular generalized-Bayesian posterior \cite{bissiri2016general}, retaining the pretrained network as the propagation prior. For residual $e$ and total scale $\varsigma^2$, the robust Student-$t$ factor \cite{lange1989robust} is
\begin{subequations}
\label{eq:student_likelihood}
\begin{align}
\ell_{\nu}(e;\varsigma^2)&\propto\left(1+\frac{e^2}{\nu\varsigma^2}\right)^{-\frac{\nu+1}{2}},
\label{eq:student_factor}\\
\varsigma^2&=\sigma^2+\sigma_{\mathrm{d}}^2,
\label{eq:student_scale}
\end{align}
\end{subequations}
where $\nu$ is the number of degrees of freedom and $\sigma_{\mathrm{d}}^2$ represents diffusion-model discrepancy. The emitter-location module updates the finite location distribution with calibrated propagation predictions. Let $\kappa_j>0$ be its measurement-dependent but position-independent likelihood power, and let $\varsigma_{\mathrm{s}}^2$ include propagation-surrogate discrepancy. The resulting emitter-location posterior is
\begin{align}
    \pi_n^{\mathrm{s}}(\tau)
    &\propto\pi_0(\tau\mid \boldsymbol{m},v)
    \prod_{j=1}^{n}
    \ell_{\nu}\!\left(
    y_j-s_{k_j}(\tau);\varsigma_{\mathrm{s}}^2
    \right)^{\kappa_j}.
    \label{eq:source_module_posterior}
\end{align}
A second likelihood power $\gamma_n\in[0,1]$ controls how strongly finite RadioDiff particles are revised by their direct residuals. The target modular posterior after $n$ measurements is
\begin{align}
    \Pi_n(x,\tau)
    &\propto q_{\theta}(x\mid \boldsymbol{m},r(\tau))
    \pi_n^{\mathrm{s}}(\tau)
    \nonumber\\
    &\quad\times\prod_{j=1}^{n}
    \ell_{\nu}(y_j-x_{k_j};\varsigma^2)^{\gamma_n}.
    \label{eq:joint_posterior}
\end{align}
The posterior in \eqref{eq:joint_posterior} uses calibrated propagation predictions to update emitter-location probabilities and RM residuals to adjust diffusion-candidate weights. Its corrected mean, projected onto known values, gives the RM estimate. The largest discrete posterior probability also gives a maximum a posteriori (MAP) emitter-location estimate, using the same location uncertainty that supports RM reconstruction.

We quantify RM and emitter-location uncertainty by the posterior expected errors of their respective estimates. Let $\pi_n(\tau)$ denote the emitter-location marginal of $\Pi_n$, let $\mu_n=\mathbb{E}_{\Pi_n}[x\mid\mathcal{D}_n]$, and let $\delta_n$ denote the MAP emitter coordinate. The RM quantity is the expected squared reconstruction error over the outdoor region, and the location quantity is the expected squared coordinate error under the finite posterior. They are defined by
\begin{align}
    R_x(\mathcal{D}_n)
    &=\mathbb{E}_{\Pi_n}
      \left[\left\|x_{\mathcal{O}}-\mu_{n,\mathcal{O}}\right\|_2^2
      \mid\mathcal{D}_n\right]
    \nonumber\\
    &=\operatorname{Tr}\!\left(\operatorname{Cov}_{\Pi_n}
      (x_{\mathcal{O}}\mid\mathcal{D}_n)\right),
    \label{eq:field_bayes_risk}\\
    R_{\tau}(\mathcal{D}_n)
    &=\sum_{\tau'\in\mathcal{H}_n}
      \pi_n(\tau')\left\|\tau'-\delta_n\right\|_2^2,
    \label{eq:source_bayes_risk}
\end{align}
Here $\mathcal{H}_n$ is the current finite emitter-location hypothesis set, with MAP decision $\delta_n=\underset{\tau'\in\mathcal{H}_n}{\operatorname{arg\,max}}\ \pi_n(\tau')$. The trace identity in \eqref{eq:field_bayes_risk} expresses expected RM reconstruction error through posterior covariance, which can be estimated from particles. The location error in \eqref{eq:source_bayes_risk} evaluates the implemented MAP coordinate estimate directly instead of using entropy as a substitute for coordinate error.

\subsection{Sequential Measurement-Location Optimization}

The sampling policy is a mapping from the collected measurements to an unmeasured outdoor location. For a budget $K$, let $\mathcal{P}=\{\mathcal{P}_n\}_{n=0}^{K-1}$ satisfy $k_{n+1}=\mathcal{P}_n(\mathcal{D}_n)$. The previously measured location set is denoted by $\mathcal{K}_n=\{k_1,\ldots,k_n\}$. The ideal design minimizes the terminal normalized RM reconstruction and emitter-location errors jointly, with expectation taken over future measurements induced by the policy. Normalization is required because RM error and coordinate error have different physical units and numerical ranges. Both initial expected errors are assumed positive. The normalized terminal objective is
\begin{align}
    J_K(\mathcal{P})
    &=\mathbb{E}_{\mathcal{P}}\!\left[
      \frac{R_x(\mathcal{D}_K)}{R_x(\mathcal{D}_0)}
      +\frac{R_{\tau}(\mathcal{D}_K)}{R_{\tau}(\mathcal{D}_0)}\right].
    \label{eq:normalized_terminal_objective}
\end{align}
The sequential problem is
\begin{subequations}\label{problem-sequential-design}
\begin{align}
    (\mathrm{P1}):\quad
    \underset{\mathcal{P}}{\operatorname{minimize}}\quad
    &J_K(\mathcal{P})
    \label{problem-sequential-design-objective}\\
    \text{s.t.}\quad
    &k_{n+1}\in\mathcal{O}\setminus\mathcal{K}_n,
      \quad 0\leq n<K,
    \label{problem-sequential-design-locations}\\
    &y_{n+1}=x_{k_{n+1}}+\epsilon_{n+1},
      \quad 0\leq n<K.
    \label{problem-sequential-design-observations}
\end{align}
\end{subequations}
Solving Problem~\ref{problem-sequential-design} exactly requires expectations over future measurement sequences and their corresponding diffusion posteriors. RMSampler approximates this terminal objective with a sequential rule based on the next measurement's expected reduction in current error. Renormalization at each round places the two remaining errors on a common scale as their magnitudes change. The RM reduction uses a moment-matched Gaussian model, while the emitter-location reduction uses a finite distribution of predicted measurements. Clipping negative predicted location improvements to zero yields the one-step criterion below, and acquired measurements subsequently support posterior correction of the reconstructed RM.

\section{Joint Active Sampling and RM Reconstruction}
\label{sec:method}

RMSampler links sequential sampling to final RM reconstruction through the measurement-updated emitter-location posterior, using separate diffusion ensembles for the two tasks, as illustrated in Fig.~\ref{fig:system_overview}. Spatial correlations from an unconditional pixel-domain ensemble support sampling, with each acquired value updating its weights and the finite location probabilities. Once sampling ends, the retained location modes condition a latent-diffusion ensemble for reconstruction. Measurement agreement then determines its posterior weights under ESS control, and projection enforces known values. Both networks retain their pretrained parameters.

\subsection{Factorized Posterior for Sampling}

The sampling posterior combines weighted background particles with finite emitter-location hypotheses. Let $\{b^{(p)},w_p\}_{p=1}^{P}$ be background particles drawn from the frozen unconditional pixel-domain RadioDiff prior, where $\sum_p w_p=1$. Let $\{\tau_m,\rho_m,\alpha_m\}_{m=1}^{M_n}$ be the current emitter-location hypotheses, their probabilities, and their fitted amplitudes. The hypotheses initially cover an outdoor grid based on the available building map, with high-probability cells refined as measurements arrive. The calibrated propagation model predicts $s_i(\tau_m;\alpha_m)$ at candidate pixel $i$. After observing $y_n$, RMSampler updates the location probabilities through the Student-$t$ factor in \eqref{eq:student_likelihood}. A small uniform probability component retains alternative locations under propagation-model mismatch. The same observation revises RM particle weights through a tempered likelihood, with regularization when their ESS becomes too small. The next sampling decision therefore uses both measurement-updated uncertainty estimates. The factorized sampling distribution $Q_n$ assigns mass $w_p\rho_m$ to each pair $(b^{(p)},\tau_m)$. Its moments guide acquisition, while $\Pi_n$ describes conditional reconstruction after sampling.

RMSampler estimates expected coordinate error using a probability-representative subset of the adaptive emitter-location grid. To retain the contribution of spatially separated hypotheses, it orders them by coordinate, forms their cumulative probability, and selects evenly spaced probability quantiles through deterministic systematic sampling. Repeated selections are combined, with their multiplicities determining the retained weights. Received-power predictions at these selected coordinates then approximate the expected error over the full spatial probability distribution.

\subsection{Normalized RM Reconstruction Uncertainty Reduction}

For candidate comparison, $x$ denotes the unobserved outdoor target values under the sampling approximation $Q_n$. Let $\Sigma_x=\operatorname{Cov}_{Q_n}(x)$, $c_k=\operatorname{Cov}_{Q_n}(x,y_k)$, and $v_k=\operatorname{Var}_{Q_n}(y_k)>0$, including sensor noise and model discrepancy. Holding these moments fixed, Gaussian conditioning gives $\Sigma_x^+=\Sigma_x-c_kc_k^{\mathsf T}/v_k$. Taking its trace shows that the predicted fractional RM error reduction is
\begin{align}
    \Delta_x(k)
    &=\frac{\left\|\operatorname{Cov}(x,y_k)\right\|_2^2}
      {\operatorname{Var}(y_k)\operatorname{Tr}(\operatorname{Cov}(x))}.
    \label{eq:field_risk_reduction}
\end{align}
In implementation, $x$ in \eqref{eq:field_risk_reduction} is restricted to a fixed-stride set of unobserved outdoor target pixels. The covariance numerator is evaluated through background, emitter-dependent, and cross-covariance Gram matrices in particle space, which avoids forming a dense $D\times D$ covariance matrix. The normalization expresses the expected error reduction as a fraction of the current RM reconstruction error for candidate $k$. A zero target trace defines this term as zero.

\subsection{Finite-Hypothesis Emitter-Location Uncertainty Reduction}

Complementing the RM term, the emitter-location component evaluates the expected coordinate error of the implemented MAP estimate. Let $d=\tau_{m^{\star}}$ with \mbox{$m^{\star}=\operatorname{arg\,max}_m\rho_m$}, and define the current expected coordinate error as $R_{\mathrm{MAP}}=\sum_m\rho_m\|\tau_m-d\|_2^2$. For candidate $k$, each retained hypothesis $i$ supplies the predicted received-power measurement $\widetilde y_{i,k}=s_k(\tau_i;\alpha_i)$. The posterior associated with that predicted measurement is
\begin{align}
    \rho_{m\mid i,k}
    &=\frac{\rho_m\ell_{\nu}
      (\widetilde y_{i,k}-s_k(\tau_m;\alpha_m);\varsigma_k^2)}
      {\sum_j\rho_j\ell_{\nu}
      (\widetilde y_{i,k}-s_k(\tau_j;\alpha_j);\varsigma_k^2)}.
    \label{eq:fantasy_source_posterior}
\end{align}
For each predicted value, recompute the MAP coordinate $d_{i,k}^+$ from \eqref{eq:fantasy_source_posterior} and set $R_{i,k}^+=\sum_m\rho_{m\mid i,k}\|\tau_m-d_{i,k}^+\|_2^2$. Averaging these conditional errors with the current hypothesis weights $\rho_i$ gives
\begin{align}
    \Delta_{\tau}(k)
    &=\max\!\left\{
      \frac{R_{\mathrm{MAP}}-\sum_i\rho_iR_{i,k}^{+}}
      {R_{\mathrm{MAP}}},0\right\}.
    \label{eq:source_risk_reduction}
\end{align}
The outer average uses a finite set of predicted measurement values, while the Student-$t$ likelihood determines each conditional location posterior. This finite approximation defines the location contribution to sampling. When the current expected coordinate error is zero, its normalized reduction is defined as zero.

\subsection{Joint Sampling Criterion and Scale Invariance}

Because the RM and emitter-location reductions are fractions of their current expected errors, both terms are dimensionless and can be combined to evaluate the predicted improvement on a common scale. RMSampler assigns candidate $k$ the sampling score
\begin{subequations}
\label{eq:joint_acquisition_score}
\begin{align}
a_k&=\Delta_x(k)+\Delta_{\tau}(k),
\label{eq:joint_score}\\
k_{n+1}&=\underset{k\in\mathcal{O}\setminus\{k_1,\ldots,k_n\}}{\operatorname{arg\,max}}\ a_k.
\label{eq:next_sampling_location}
\end{align}
\end{subequations}
\begin{theorem}[One-step joint estimation-error minimization]
\label{thm:joint_one_step_risk}
Let $\Sigma_x=\operatorname{Cov}(x)$, $c_k=\operatorname{Cov}(x,y_k)$, $v_k=\operatorname{Var}(y_k)>0$, and $R_x=\operatorname{Tr}(\Sigma_x)>0$. Let $R_{\tau}=R_{\mathrm{MAP}}>0$, and define the post-measurement RM reconstruction error and expected emitter-location error by
\begin{subequations}
\label{eq:post_measurement_joint_risks}
\begin{align}
R_x^{+}(k)&=\operatorname{Tr}\!\left(\Sigma_x-\frac{c_kc_k^{\mathsf{T}}}{v_k}\right),
\label{eq:post_measurement_rm_error}\\
\overline{R}_{\tau}^{+}(k)&=\sum_i\rho_iR_{i,k}^{+}.
\label{eq:post_measurement_location_error}
\end{align}
\end{subequations}
The RMSampler sampling score satisfies
\begin{align}
    a_k
    &=2-\left[
      \frac{R_x^{+}(k)}{R_x}
      +\min\!\left\{
      \frac{\overline{R}_{\tau}^{+}(k)}{R_{\tau}},1
      \right\}
      \right].
    \label{eq:joint_score_risk_equivalence}
\end{align}
Hence, maximizing $a_k$ exactly minimizes the clipped normalized one-step joint estimation error under the Gaussian RM approximation and the finite emitter-location measurement model with Student-$t$ likelihoods.
\end{theorem}

\begin{radioproof}
Taking the trace of the Gaussian rank-one covariance update gives $\Delta_x(k)=1-R_x^+(k)/R_x$. The clipped coordinate reduction equals $\Delta_\tau(k)=1-\min\{\overline R_\tau^+(k)/R_\tau,1\}$. Adding these identities yields \eqref{eq:joint_score_risk_equivalence}. Since the additive constant is independent of $k$, maximizing the score minimizes the stated one-step surrogate.
\end{radioproof}

Theorem~\ref{thm:joint_one_step_risk} connects candidate ranking to the joint estimation objective under the stated Gaussian and finite-hypothesis models. Recomputing both terms after each measurement adapts the next selection to the acquired information. Increasing the budget continues the same sequence, giving nested measurement sets for cross-budget comparison.

\begin{corollary}[Affine representation invariance]
\label{cor:affine_invariance}
Apply $z'=cz+d$ with $c\neq0$ to every RM value, observed or predicted measurement, and power prediction, and multiply every sensor and model-discrepancy variance by $c^2$. Then $\Delta_x'(k)=\Delta_x(k)$, $\Delta_{\tau}'(k)=\Delta_{\tau}(k)$, and $a_k'=a_k$ for every candidate. The maximizing location set in \eqref{eq:joint_acquisition_score} is therefore unchanged.
\end{corollary}

\begin{radioproof}
Translations vanish from covariance and from residuals, whereas multiplication by $c$ gives
\begin{subequations}
\label{eq:affine_power_transform}
\begin{align}
    \operatorname{Cov}(x',y_k')
    &=c^2\operatorname{Cov}(x,y_k),
    \label{eq:affine_covariance}\\
    \operatorname{Var}(y_k')
    &=c^2\operatorname{Var}(y_k),
    \label{eq:affine_measurement_variance}\\
    \operatorname{Tr}(\operatorname{Cov}(x'))
    &=c^2\operatorname{Tr}(\operatorname{Cov}(x)),
    \label{eq:affine_target_trace}\\
    \frac{(\widetilde y_{i,k}'-s_k'(\tau_m;\alpha_m))^2}
      {\varsigma_k'^2}
    &=\frac{(\widetilde y_{i,k}-s_k(\tau_m;\alpha_m))^2}
      {\varsigma_k^2}.
    \label{eq:affine_standardized_residual}
\end{align}
\end{subequations}
The RM error-reduction numerator and denominator in \eqref{eq:field_risk_reduction} therefore acquire the same factor $c^4$. Equation~\eqref{eq:affine_standardized_residual} leaves every relative Student-$t$ factor unchanged, so the posteriors conditioned on predicted measurements, their MAP estimates, and their expected coordinate errors are identical. Hence, $\Delta_x'(k)=\Delta_x(k)$, $\Delta_{\tau}'(k)=\Delta_{\tau}(k)$, and $a_k'=a_k$, preserving every pairwise ordering and tie among candidate locations.
\end{radioproof}

Corollary~\ref{cor:affine_invariance} shows that affine received-power conversions preserve the selected sampling locations when sensor and model-discrepancy variances are transformed consistently.

\subsection{Posterior-Corrected Conditional Diffusion}

Once sampling ends, the updated emitter-location posterior determines the conditions for final RM reconstruction. RMSampler groups that posterior into separated modes. For mode $m$, the coordinate $\tau_m$ is rasterized as $r(\tau_m)$ and supplied to the pretrained conditional latent-diffusion model of RadioDiff. The conditioning inputs follow the model's original training specification and contain only the building mask and measurement-derived emitter-location raster. Acquired outdoor received power enters the likelihood and projection operations below, while the pretrained model's parameters and architecture remain fixed.

For these retained emitter-location modes, candidate RMs are generated from independent latent noise. Let $L$ denote the number of RM particles per emitter-location mode. Repeating each mode condition $L$ times produces the joint ensemble,
\begin{align}
    x^{(m,\ell)}
    &=G_{\theta}\!\left(z^{(m,\ell)};\boldsymbol{m},r(\tau_m)\right),
    \label{eq:batched_diffusion_particles}\\
    m&=1,\ldots,M_{\star},
    \qquad \ell=1,\ldots,L.
    \nonumber
\end{align}
The prior weight of particle $(m,\ell)$ is $u_{m,\ell}^{0}=\rho_m/L$. Memory-bounded batches evaluate these reverse-diffusion trajectories with global posterior normalization, retaining the specified prior weights and independent latent noise. The ensemble represents RM variation within each mode and emitter-location uncertainty across modes.

To favor candidate RMs that explain the collected measurements, RMSampler uses a robust negative log-likelihood, defined for particle $(m,\ell)$ by
\begin{align}
    E_{m,\ell}
    &=\frac{\nu+1}{2}\sum_{j=1}^{K}
      \log\!\left(
      1+\frac{(y_j-x_{k_j}^{(m,\ell)})^2}
      {\nu(\sigma^2+\sigma_{\mathrm{d}}^2)}
      \right),
    \label{eq:particle_robust_energy}\\
    u_{m,\ell}(\gamma)
    &=\frac{u_{m,\ell}^{0}\exp(-\gamma E_{m,\ell})}
      {\sum_{a,b}u_{a,b}^{0}\exp(-\gamma E_{a,b})}.
    \label{eq:tempered_particle_weights}
\end{align}
The Student-$t$ loss limits the influence of measurements poorly explained by the finite ensemble, with sensor noise and diffusion-model discrepancy included in the residual scale. The resulting weights in \eqref{eq:tempered_particle_weights} favor candidates that better match acquired values, so the measurements also affect estimates at unsampled locations.

ESS-guided correction balances measurement fit against weight concentration. Define \mbox{$\operatorname{ESS}(\gamma)=(\sum_{m,\ell}u_{m,\ell}(\gamma)^2)^{-1}$} \cite{kong1994sequential}, with nominal target $N_{\min}$. The likelihood power is selected by
\begin{align}
    \gamma^\star&=
    \begin{cases}
    1, & \operatorname{ESS}(1)\geq N_{\min},\\
    \gamma_{\rm lo}, & \text{otherwise},
    \end{cases}
    \label{eq:ess_tempering_rule}
\end{align}
where $\gamma_{\rm lo}$ is the lower endpoint after 40 bracket-search steps initialized at $0$ and $1$. An ESS-admissible midpoint replaces the lower endpoint, and any other midpoint replaces the upper endpoint. If no admissible midpoint is found, the endpoint remains zero and retains the mode-weighted particle prior. Unequal mode probabilities can make its initial ESS lower than the nominal target. The following result characterizes measurement fitting for every selected likelihood power.

\begin{theorem}[Variational posterior correction]
\label{thm:variational_tempering}
Index the joint particles by $r\in\mathcal I$. Assume $u_r^0>0$, $\sum_r u_r^0=1$, and finite losses $E_r$. For each $\gamma\in[0,1]$, the weights in \eqref{eq:tempered_particle_weights} uniquely solve the KL-regularized problem
\begin{align}
u(\gamma)&=\underset{q\in\Delta_{\mathcal I}}{\operatorname{arg\,min}}
\left\{\operatorname{KL}(q\Vert u^0)+\gamma\sum_rq_rE_r\right\},
\label{eq:variational_particle_posterior}
\end{align}
where $\Delta_{\mathcal I}$ is the probability simplex. Their expected robust loss $\overline E(\gamma)=\sum_ru_r(\gamma)E_r$ satisfies
\begin{align}
\frac{\mathrm d\overline E(\gamma)}{\mathrm d\gamma}
&=-\operatorname{Var}_{u(\gamma)}(E)\leq0.
\label{eq:tempering_energy_monotonicity}
\end{align}
\end{theorem}

\begin{radioproof}
The simplex-constrained first-order condition gives $q_r\propto u_r^0\exp(-\gamma E_r)$. Normalization yields \eqref{eq:tempered_particle_weights}, and strict convexity of the KL term gives uniqueness. Differentiating the normalized weights gives $\mathrm du_r/\mathrm d\gamma=u_r(\overline E-E_r)$. Multiplying by $E_r$ and summing gives \eqref{eq:tempering_energy_monotonicity}.
\end{radioproof}

Theorem~\ref{thm:variational_tempering} explains how increasing likelihood power reduces expected robust loss for the fixed ensemble. ESS is evaluated directly during the numerical search because unequal prior weights can make it nonmonotone in that power. The search and its zero-correction fallback use acquired measurements only.

\subsection{Posterior-Mean Reconstruction and Emitter Localization}

The corrected particle weights determine both the RM and the emitter-location estimate. Their weighted mean gives the unprojected RM
\begin{align}
    \overline{x}
    &=\sum_{m=1}^{M_{\star}}\sum_{\ell=1}^{L}
      u_{m,\ell}(\gamma^{\star})x^{(m,\ell)}.
    \label{eq:posterior_mean_map}
\end{align}
Summing corrected particle weights within each mode yields the emitter-location probabilities $\overline{\rho}_m=\sum_{\ell=1}^{L}u_{m,\ell}(\gamma^{\star})$. The resulting MAP estimate $\widehat{\tau}=\tau_{\operatorname{arg\,max}_m\overline{\rho}_m}$ uses the same measurement information as the RM estimate.

Projection imposes the bounded measurement and building constraints after posterior weighting. Let $\mathcal K=\{k_1,\ldots,k_K\}$ and define $y_{\rm b}(i)=\operatorname{clip}(y(i),0,1)$ at each acquired pixel. The likelihood uses the raw noisy value $y(i)$, while the final bounded RM uses
\begin{align}
    \widehat{x}_i
    &=
    \begin{cases}
        y_{\rm b}(i), & i\in\mathcal{K},\\
        v_i, & i\in\mathcal{B},\\
        \min\{1,\max\{0,\overline{x}_i\}\}, & \text{otherwise}.
    \end{cases}
    \label{eq:exact_projection}
\end{align}
Equation~\eqref{eq:exact_projection} matches the bounded measurement targets and prescribed building values. Unsampled pixels retain the posterior estimate clipped to the normalized RM range. Weight estimation precedes projection, so enforced values do not enter the particle likelihood.

\subsection{Projection and Computational Properties}

The final projection imposes measurement and building constraints with the minimum squared change to the posterior mean. These constraints define the feasible RM set
\begin{align}
    \mathcal{C}_{K}
    &=\left\{z\in\mathbb{R}^{D}:
      z_i=y_{\rm b}(i),\ i\in\mathcal{K};\ 
      z_i=v_i,\ i\in\mathcal{B};\right.
    \nonumber\\
    &\left.\hspace{45mm}
      0\leq z_i\leq1,\ i\notin\mathcal{K}\cup\mathcal{B}
      \right\}.
    \label{eq:feasible_rm_set}
\end{align}
This set fixes bounded measurement targets and building values while constraining unobserved outdoor values to the normalized RM range. The projection in \eqref{eq:exact_projection} applies these constraints coordinatewise, leading to the following property.

\begin{proposition}[Minimum-Distortion Projection]
\label{prop:projection_readout}
Let $\mathcal{P}_{K}$ denote the operator in \eqref{eq:exact_projection}. For every posterior mean $p\in\mathbb{R}^{D}$, its output is the unique solution
\begin{align}
    \mathcal{P}_{K}(p)
    &=\underset{z\in\mathcal{C}_{K}}{\operatorname{arg\,min}}
      \ \lVert z-p\rVert_2^2.
    \label{eq:projection_variational_form}
\end{align}
Moreover, the projection is nonexpansive for any $p,q\in\mathbb{R}^{D}$, satisfying $\lVert\mathcal{P}_{K}(p)-\mathcal{P}_{K}(q)\rVert_2\leq\lVert p-q\rVert_2$.
\end{proposition}

\begin{radioproof}
The feasible set in \eqref{eq:feasible_rm_set} is a Cartesian product of singleton coordinates and closed intervals. It is therefore nonempty, closed, and convex whenever the measured and building index sets are disjoint, as ensured by $\mathcal{K}\subseteq\mathcal{O}$. The squared Euclidean objective separates across coordinates and is strictly convex. A constrained coordinate is uniquely fixed to $y_{\rm b}(i)$ or $v_i$, while every remaining coordinate has the scalar minimizer $\min\{1,\max\{0,p_i\}\}$. These coordinate solutions reproduce \eqref{eq:exact_projection} and establish \eqref{eq:projection_variational_form}. Fixed coordinates contribute zero to the difference between two projected RMs, and scalar clipping onto $[0,1]$ is one-Lipschitz. Summing the coordinatewise squared inequalities establishes nonexpansiveness.
\end{radioproof}

Proposition~\ref{prop:projection_readout} gives the minimum-distortion RM satisfying the bounded measurement and building constraints. For a fixed constraint set, nonexpansiveness prevents projection from amplifying differences between posterior means.

The corrected posterior can be evaluated in batches using one global weight normalization. Relabel the $N=M_{\star}L$ joint particles by $r\in\mathcal{I}$, collect them as $X=[x^{(1)},\ldots,x^{(N)}]$, and define $h_r=\log u_r^0-\gamma^{\star}E_r$. The posterior mean and mode masses can then be evaluated through one global normalization,
\begin{subequations}
\label{eq:global_batched_readout}
\begin{align}
u_r&=\frac{\exp(h_r-h_{\max})}{\sum_j\exp(h_j-h_{\max})},
\label{eq:normalized_particle_weight}\\
h_{\max}&=\max_j h_j,
\label{eq:log_weight_maximum}\\
\overline{x}&=Xu,
\label{eq:batched_rm_mean}\\
\overline{\rho}_m&=\sum_{r:\,\tau(r)=\tau_m}u_r.
\label{eq:batched_mode_mass}
\end{align}
\end{subequations}
This expression permits the diffusion generator, robust losses, and posterior accumulation to use different batch sizes without changing the estimator.

\begin{proposition}[Batch and chunk invariance]
\label{prop:batch_chunk_invariance}
Fix the latent noises, emitter-location conditions, energies, and prior weights of the joint particle ensemble. Partition $\mathcal{I}$ into arbitrary disjoint chunks $\{\mathcal{I}_c\}_{c=1}^{C}$. If all chunks use the global normalizer in \eqref{eq:global_batched_readout}, then chunked generation and accumulation produce exactly the same $\overline{x}$, $\overline{\rho}_m$, and $\widehat{\tau}$ as an unchunked batch.
\end{proposition}

\begin{radioproof}
The fixed diffusion generator acts samplewise once the latent noise and condition of each particle are fixed, so batching changes only execution order. Let \mbox{$a_r=\exp(h_r-h_{\max})$} and $A=\sum_c\sum_{r\in\mathcal{I}_c}a_r$. Disjointness of the chunks gives
\begin{subequations}
\label{eq:chunked_posterior_accumulation}
\begin{align}
\overline{x}&=\frac{1}{A}\sum_{c=1}^{C}\sum_{r\in\mathcal{I}_c}a_rx^{(r)},
\label{eq:chunked_rm_mean}\\
\overline{\rho}_m&=\frac{1}{A}\sum_{c=1}^{C}\sum_{\substack{r\in\mathcal{I}_c\\\tau(r)=\tau_m}}a_r.
\label{eq:chunked_mode_mass}
\end{align}
\end{subequations}
Both expressions are regroupings of the unchunked sums in \eqref{eq:global_batched_readout}. Their values and the resulting highest-probability emitter-location mode are therefore unchanged.
\end{radioproof}

Proposition~\ref{prop:batch_chunk_invariance} shows that splitting a fixed candidate ensemble into chunks preserves the reconstructed RM and emitter-location estimate when global weight normalization is retained. A chunk of $B$ RMs requires $\mathcal{O}(BD)$ storage instead of $\mathcal{O}(ND)$. The arithmetic for $N$ reverse trajectories is unchanged, and batching permits parallel evaluation across particles and location modes. Loss evaluation uses only the $K$ measured entries of each RM, after which its contribution to the posterior mean can be accumulated online. The implementation can therefore select batch sizes according to available memory without changing the defined posterior calculation.

Particle-space computation also limits sampling memory by avoiding dense covariance storage across RM pixels. With $T$ RM target pixels, $C$ candidates, $P$ background particles, and $S$ retained emitter-location hypotheses, the Gram computation scales as $\mathcal{O}((T+C)(P+S)^2)$, while evaluating predicted measurements under emitter-location hypotheses adds $\mathcal{O}(CS^2)$. Candidate chunks limit prediction storage without changing the result, making computation depend on particle and hypothesis counts rather than a dense pixelwise covariance matrix.

\subsection{Adaptation to Real-World Measured Data}

For real-world measured data, RM reconstruction supplements the frozen prior with fixed received-power normalization and measurement-based residual correction. RMSampler aggregates repeated measurements within each visited pixel by their median and converts received power in decibels relative to one milliwatt (dBm) through the fixed transform $\widetilde y=\operatorname{clip}((y_{\rm dBm}+147)/100,0,1)$. Receiver coordinates define the reconstruction region, with Tx coordinates excluded from crop and feature construction. Let $\bar x$ be the measurement-weighted mean of diffusion particles and let $z_i$ collect terrain, building, and local-density features available at inference. A ridge-regularized environment kernel fits residuals $y_k-\beta\bar x_k$ at sampled locations and predicts the correction at unmeasured receiver locations, giving
\begin{align}
    \widehat x_i^{\rm meas}
    =\Pi_{[0,1]}\!\left(\beta\bar x_i+\eta\widehat r_i(z_i,\mathcal D_n)\right),
    \label{eq:measured_residual_adapter}
\end{align}
where $\widehat r_i$ is fitted to $\mathcal D_n$. The positive $\beta$ preserves the diffusion-prior contribution, while the fitted term accounts for received-power discrepancies associated with the observed environment.

The same simulation-to-measurement mismatch can also affect sampling, so spatial coverage supplements uncertainty-based selection when the transferred covariance does not adequately describe the target environment. We combine the uncertainty score with normalized maximin coverage $c_k^{\rm cov}$ through
\begin{align}
    a_k^{\rm meas}&=(1-q)c_k^{\rm cov}+q\,\mathcal{N}_n\!\left(\Delta_x^{\rm env}+\omega_\tau\Delta_\tau\right)(k),
    \label{eq:measured_acquisition}
\end{align}
Here, $\Delta_x^{\rm env}(k)$ is the mean variance explained by the sampled locations together with candidate $k$ under the environment kernel. The operator $\mathcal{N}_n$ rescales a score vector to $[0,1]$ using its minimum and maximum over the currently unmeasured candidates, and returns zero when all candidate scores coincide. The coverage values are rescaled in the same way to obtain $c_k^{\rm cov}$. The emitter-location probabilities are updated after each new measurement, and development and validation cells fix $(\beta,\eta,q,\omega_\tau)$ before testing. The measured-data input follows the available environmental information, using an all-free building mask with terrain descriptors for Chongqing when a compatible building raster is unavailable.

\section{Experimental Evaluation}
\label{sec:experiments}

The evaluation isolates sampling gains by applying the same conditional diffusion reconstructor to RMSampler, Uniform, and Random on simulated RMs. External baselines retain their complete pipelines. On real-world measured data, common-location and common-reconstructor controls separate reconstruction quality from sampling effectiveness.

\subsection{Protocol, Baselines, and Metrics}

The simulation study uses dominant-path-model RMs from RadioMapSeer \cite{levie2021radiounet}. A fixed scene-level permutation with seed 42 partitions the data into 500 training, 100 validation, and 100 test scenarios, with no scene overlap between partitions. We evaluate all 100 test scenarios, which are unseen during training, using frozen diffusion models for zero-shot RM reconstruction. Each scenario provides 80 Tx positions with a separate single-Tx RM for each position. We use the first provided Tx configuration in each test scenario, giving one $256\times256$ RM per scenario with only that Tx's contribution. The exact building map is available at inference as stipulated in Section~\ref{sec:system_model}, while the true Tx coordinate and unmeasured outdoor received-power values are reserved for evaluation. We evaluate 8 and 16 outdoor measurements, denoted by B8 and B16, with normalized additive-noise standard deviation $0.02$. All methods share the test scenarios, budgets, masks, and noise convention. Known building-pixel values are excluded from the sampling budget.

The three sampling strategies use the same conditional reconstruction settings, with measurement sets generated using seed 42031. An unconditional pixel-domain diffusion model supplies 12 sampling particles with 16 reverse steps. The global RM criterion uses a four-pixel target stride and at most 96 representative emitter-location hypotheses. Final reconstruction generates four conditional latent-diffusion particles per retained Tx mode, with three reverse steps and at most four modes. Each mode uses a peak-normalized Gaussian raster with one-pixel standard deviation. The Student-$t$ likelihood has four degrees of freedom, discrepancy standard deviation $0.1$, and a nominal ESS target equal to the larger of two particles and $0.35$ times the ensemble size. Posterior correction and projection use each strategy's acquired measurements and fixed scene-specific latent-noise streams. Conditional inference uses an NVIDIA RTX~3060~Ti, four-particle batches per mode, and global weight normalization. External methods are evaluated on an RTX~4090. All neural parameters and architectures remain fixed.

The external baselines are independent paper-constrained reimplementations that retain published sampling and reconstruction mechanisms with the dataset and budget adaptations specified here. Ensemble GP active learning (EGP-AL) uses 11 radial-basis random-feature GP experts, predictive-likelihood weighting, Gaussian-mixture variance (GM-Var) sampling, and mixture-mean reconstruction~\cite{polyzos2024bayesian}. Its initialization is adapted to the sparse budget, whereas the published experiment starts from 100 labeled points. GP-UCB-CE retains online GP fitting, the published upper-confidence reward, constrained unicycle motion, a five-step horizon, and gradient-enhanced cross-entropy planning~\cite{zhang2025adaptive}.

The geometry and low-rank baselines use different structural information for sampling and reconstruction. GACP-AE retains its $32\times32$ lattice, geometry-aware covariance, pivoted Cholesky projection, and published autoencoder~\cite{chai2026geometry}. The geometry uses the documented building-footprint adaptation in place of unavailable obstacle heights. After 100 training epochs, the model reaches validation mean squared error $0.00407$ without projection onto the measured values. EMLS with matrix completion (EMLS-MC) retains its two-round sampler and uses singular value thresholding (SVT) for reconstruction~\cite{sun2024energy}. Its published protocol spans $5\%$ to $45\%$ sampling, while B8 and B16 expose roughly $0.01\%$ and $0.02\%$ of the grid. After the fixed 600 iterations, its observed-entry residuals remain $0.6075$ and $0.6099$, exceeding the unchanged $10^{-4}$ threshold. Table~\ref{tab:main_formal_results} therefore records non-convergence (NC) under this stopping rule. External methods use their own sampling and reconstruction pipelines, alongside the stated shared-reconstruction controls. Unmeasured received-power values remain excluded from every method's inference inputs.

Simulation reports outdoor NMSE and SSIM~\cite{wang2004ssim}, converting NMSE to decibels per scenario before averaging. Peak ratio compares predicted and reference maximum normalized RM values within five pixels of the true emitter, with an ideal value of one. Top-1\% recall is the fraction of reference top-1\% outdoor pixels also identified among the predicted top-1\% pixels. For localization, the three sampling policies use the MAP estimate from corrected Tx-mode probabilities, while external methods use the outdoor RM peak. Errors measure Euclidean distance to true antenna coordinates, with hit rates within 5, 10, and 20 pixels. Means cover the same 100 scenarios. Figures use 5000 scene-bootstrap resamples for mean intervals~\cite{efron1979bootstrap}, while paired differences use 20000 resamples and 100000 sign flips. Holm adjustment covers both budgets and the two sampling comparators for each metric, with external-pipeline comparisons forming a separate family. B8 and B16 expose less than $0.025\%$ of the grid, so evaluation includes far more unsampled than measured locations.

\begin{table*}[!t]
\centering
\caption{RM reconstruction on 100 held-out RadioMapSeer scenarios. RMSampler, Uniform, and Random share the conditional diffusion reconstructor. External methods retain their complete pipelines. Bold and underlined values mark the best method and best baseline using unrounded means. NC denotes an unmet SVT stopping criterion at B8/B16.}
\label{tab:main_formal_results}
\setlength{\tabcolsep}{3.8pt}
\renewcommand{\arraystretch}{1.12}
\resizebox{\textwidth}{!}{%
\begin{tabular}{@{}lccccc|ccccc@{}}
\toprule
& \multicolumn{5}{c|}{8 measurements} & \multicolumn{5}{c}{16 measurements} \\
\cmidrule(lr){2-6}\cmidrule(lr){7-11}
Method & NMSE / dB $\downarrow$ & SSIM $\uparrow$ & Peak ratio $\uparrow$ & Top-1\% recall $\uparrow$ & Tx error / px $\downarrow$ & NMSE / dB $\downarrow$ & SSIM $\uparrow$ & Peak ratio $\uparrow$ & Top-1\% recall $\uparrow$ & Tx error / px $\downarrow$ \\
\midrule
\rowcolor{blue!7}
RMSampler & \textbf{-18.1546} & \textbf{0.9178} & \textbf{0.7415} & \textbf{0.4416} & \textbf{20.12} & \textbf{-19.2801} & \textbf{0.9255} & \textbf{0.8092} & \textbf{0.5265} & \textbf{14.69} \\
Uniform & -14.4682 & 0.8788 & 0.6020 & 0.2045 & 39.60 & -17.0566 & 0.9009 & 0.7065 & 0.3685 & 25.54 \\
Random & \underline{-16.2121} & \underline{0.8977} & \underline{0.6721} & \underline{0.3062} & \underline{31.24} & \underline{-18.2314} & \underline{0.9180} & \underline{0.7534} & \underline{0.4237} & \underline{19.48} \\
EGP-AL (GM-Var) & -5.5441 & 0.5199 & 0.2905 & 0.0470 & 71.50 & -8.6235 & 0.6470 & 0.4307 & 0.0860 & 56.28 \\
GP-UCB-CE & -6.3925 & 0.7221 & 0.3214 & 0.0258 & 108.34 & -6.5208 & 0.7310 & 0.3617 & 0.0407 & 91.11 \\
GACP-AE & -3.0874 & 0.3960 & 0.1836 & 0.0462 & 75.78 & -3.5133 & 0.3970 & 0.2458 & 0.0695 & 62.08 \\
EMLS-MC$^{\dagger}$ & \multicolumn{10}{c}{\textcolor{gray!75!black}{NC because the SVT residual remained above the fixed $10^{-4}$ threshold at B8/B16}} \\
\midrule
\rowcolor{gray!12}
RMSampler gain & +1.94 dB & +0.0201 & +0.0694 & +0.1353 & +11.12 px & +1.05 dB & +0.0074 & +0.0558 & +0.1029 & +4.79 px \\
\bottomrule
\end{tabular}
}
\end{table*}

\begin{figure}[!t]
\centering
\includegraphics[width=\columnwidth]{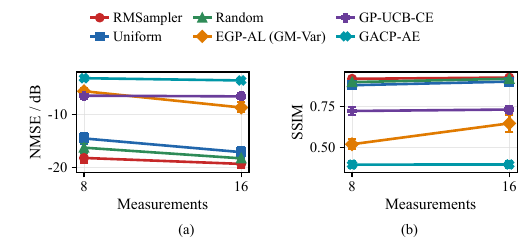}
\caption{Mean outdoor NMSE and SSIM with scenario-bootstrap 95\% confidence intervals.}
\label{fig:budget_quality}
\end{figure}

\begin{figure}[!t]
\centering
\includegraphics[width=\columnwidth]{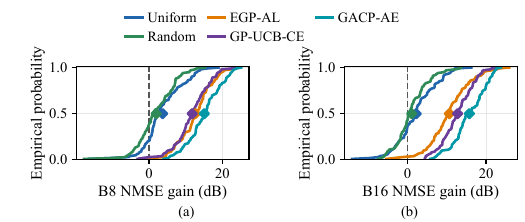}
\caption{Paired RMSampler NMSE gains over baselines with reported reconstruction results.}
\label{fig:paired_nmse_gain}
\end{figure}

\subsection{Aggregate RM Recovery}

With the conditional reconstructor fixed, Table~\ref{tab:main_formal_results} isolates the benefit of sampling-location selection against Uniform and Random. RMSampler achieves NMSE of $-18.1546$ and $-19.2801$~dB at B8 and B16, improving on Random by $1.94$ and $1.05$~dB. It also has the highest mean SSIM at both budgets. The gain row uses the strongest reported baseline for each metric. Figure~\ref{fig:budget_quality} shows mean uncertainty intervals, and Fig.~\ref{fig:paired_nmse_gain} shows the paired scenario-level distributions. Outdoor metrics include all unsampled locations, beyond the pixels fixed by measurement projection.

\begin{figure}[!t]
\centering
\includegraphics[width=\columnwidth]{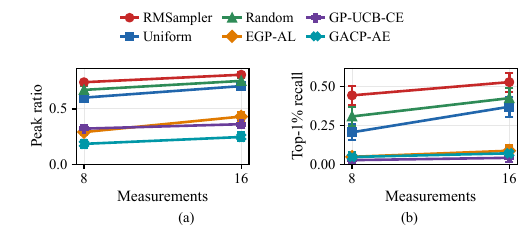}
\caption{Peak-amplitude recovery and top-1\% outdoor-pixel recall.}
\label{fig:peak_recovery}
\end{figure}

\begin{figure*}[!t]
\centering
\includegraphics[width=0.85\textwidth]{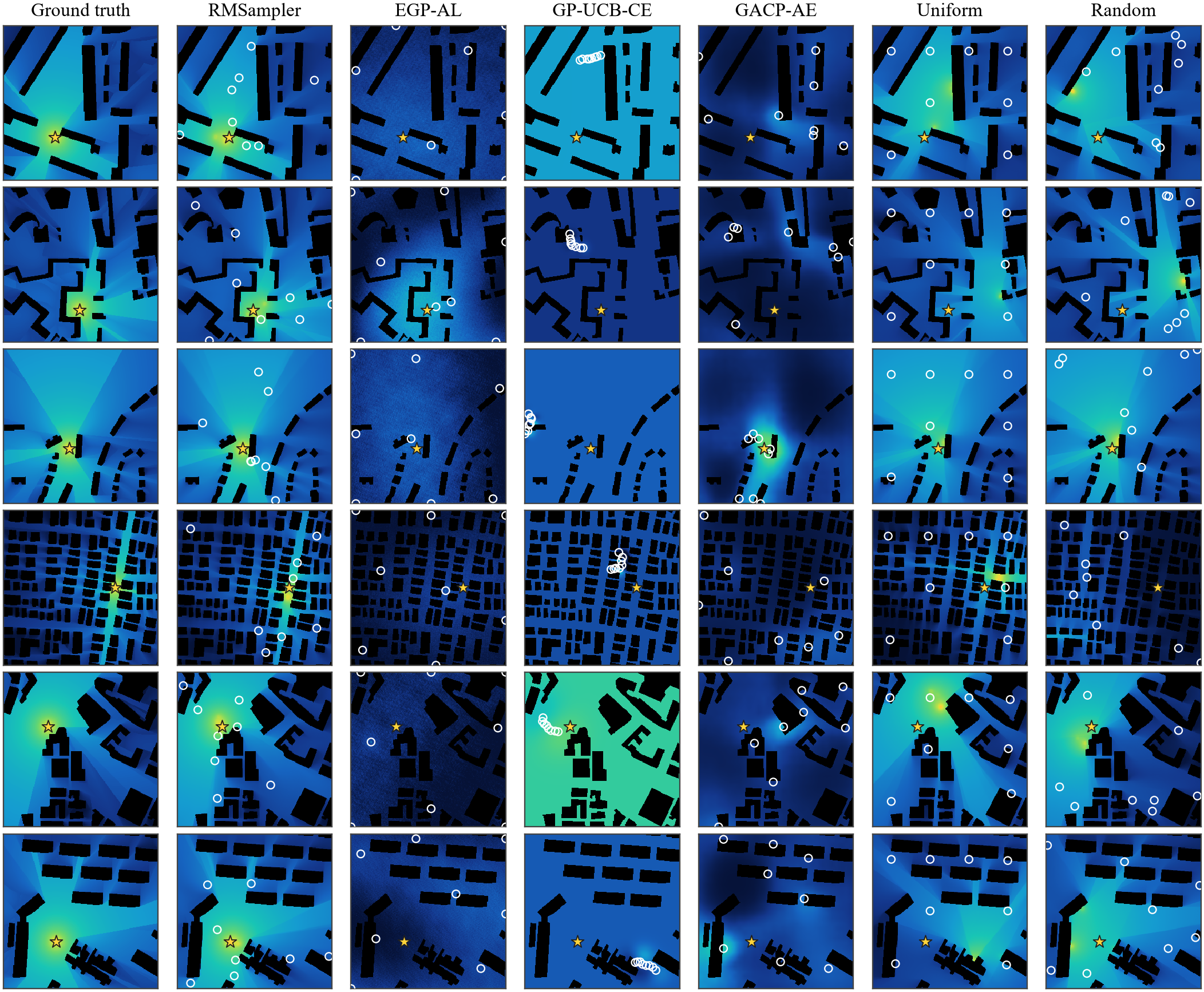}
\caption{B8 RM comparison on six test scenarios selected by the largest minimum NMSE gain over every compared baseline. Columns are methods, rows are scenarios, circles are sampling locations, and stars indicate true Tx locations.}
\label{fig:qualitative_gallery}
\end{figure*}

\subsection{Spatial Fidelity and Visual RM Comparison}

High-power-region metrics assess recovery of the strongest interference regions. At B8, RMSampler raises peak ratio from Random's $0.6721$ to $0.7415$ and top-1\% recall from $0.3062$ to $0.4416$ under the same conditional reconstructor. Both metrics improve further at B16 in Fig.~\ref{fig:peak_recovery}, showing that the reconstruction-error reduction is accompanied by improved recovery of strong-signal regions.

Figure~\ref{fig:qualitative_gallery} compares RMs on a common normalized color scale. The six scenarios are selected from the test results by the largest minimum B8 NMSE gain over the compared baselines, illustrating favorable cases. Their reconstructions show closer agreement in high-power regions, street-aligned propagation, and building shadows. Complete-cohort metrics provide the aggregate comparison.

\begin{figure}[!t]
\centering
\includegraphics[width=\columnwidth]{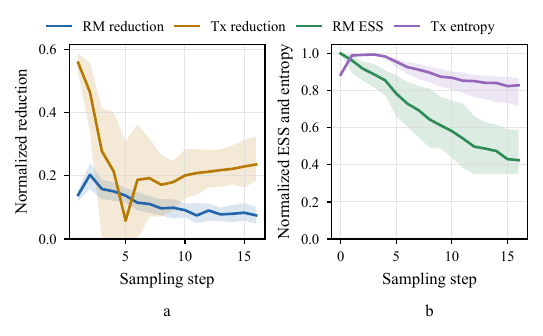}
\caption{Median normalized expected-error reductions and posterior concentration, with shaded interquartile ranges.}
\label{fig:acquisition_dynamics}
\end{figure}

\subsection{Measurement-Location Behavior}

The sampling criterion changes its balance as acquired measurements resolve emitter-location ambiguity. Figure~\ref{fig:acquisition_dynamics} shows a larger emitter-location term early in sampling, followed by a sustained RM reconstruction term. Particle ESS and location entropy describe the associated weight concentration and remaining ambiguity. Together with Fig.~\ref{fig:motivation_locations}, these traces explain how the measurement updates guide subsequent sampling locations.

\subsection{Statistical Reliability}

Paired inference confirms sampling gains with the conditional reconstructor held fixed. Against Random, mean NMSE gains are $1.943$ and $1.049$~dB at B8 and B16, with 95\% intervals of $[0.900,2.972]$ and $[0.223,1.909]$~dB and Holm-adjusted $p=0.00086$ and $0.0170$. Uniform comparisons yield $3.686$ and $2.223$~dB, with positive intervals and adjusted $p<0.001$. Both SSIM comparisons also pass adjusted testing at both budgets. Figure~\ref{fig:paired_nmse_gain} retains the complete distributions, with RMSampler having lower NMSE than Random in 62 and 57 of the 100 scenarios. The three external-pipeline NMSE comparisons have positive intervals at both budgets. Per-scenario predictions, measurement sets, and non-converged outcomes are retained for reproduction.

\subsection{Emitter Localization Performance}

The same corrected Tx posterior supports a common localization comparison across the three sampling strategies. At B16, mean location error decreases from Random's $19.48$ to $14.69$ pixels in Table~\ref{tab:main_formal_results}. The paired reduction is $4.79$ pixels, with a 95\% interval of $[0.47,9.49]$ pixels and Holm-adjusted $p=0.0345$. All four coordinate-error comparisons against Uniform and Random pass adjusted testing. RMSampler hit rates within 5, 10, and 20 pixels are $0.16$, $0.40$, and $0.68$ at B8 and $0.25$, $0.51$, and $0.78$ at B16.

\subsection{Real-World Measured-Data Validation}

All RadioMapSeer-pretrained neural parameters remain fixed during the Vienna and Chongqing experiments, with no fine-tuning on measured data. Recorded reference signal received power (RSRP) comes from Vienna scanner measurements~\cite{wiedner2026vienna} and Chongqing ground measurements~\cite{xu2026complex}. The strongest synchronization signal block (SSB) per snapshot is median-aggregated in 4-m pixels within the densest 1-km crop of the largest receiver trajectory component. Vienna supplies 1-m building/terrain maps, while Chongqing uses terrain and an all-free mask. Normalization is fixed to $[-147,-47]$~dBm. Development, validation, and test splits contain four, four, and four cells for Vienna and four, four, and eight for Chongqing. Three fixed seeds are averaged per cell. True emitter coordinates and unacquired values are excluded. Development and validation select numerical settings, while acquired measurements update only posterior quantities and residual coefficients. RMSE and 90th percentile absolute error (P90) are evaluated at held-out receiver locations, with cells as statistical units.

\begin{table}[!t]
\centering
\caption{RMSE in dB on held-out real-world measurements with identical sampling locations and frozen network parameters.}
\label{tab:measured_transfer}
\setlength{\tabcolsep}{3.0pt}
\renewcommand{\arraystretch}{1.02}
\resizebox{\columnwidth}{!}{%
\begin{tabular}{@{}lrrrr@{}}
\toprule
& \multicolumn{2}{c}{Vienna RMSE / dB} & \multicolumn{2}{c}{Chongqing RMSE / dB} \\
\cmidrule(lr){2-3}\cmidrule(lr){4-5}
Method & B8 & B16 & B8 & B16 \\
\midrule
\rowcolor{blue!7}RMSampler & \textbf{4.84} & \textbf{4.29} & \textbf{3.75} & \textbf{3.08} \\
Kriging & 5.74 & \underline{4.68} & \underline{3.92} & \underline{3.17} \\
GACP-AE & \underline{4.97} & 5.03 & 5.22 & 5.52 \\
RMSampler posterior & 7.85 & 6.92 & 5.24 & 5.37 \\
RadioDiff prior & 9.17 & 9.22 & 6.12 & 6.30 \\
Saipp-Net & 9.21 & 8.97 & 10.30 & 9.68 \\
Physics-Feature U-Net & 9.47 & 8.43 & 13.18 & 12.39 \\
Sparse-Guided RadioUNet & 10.97 & 9.85 & 9.79 & 11.03 \\
RadioUNet & 13.48 & 12.01 & 16.15 & 16.31 \\
SIP2Net & 13.71 & 11.30 & 8.41 & 7.55 \\
\bottomrule
\end{tabular}
}
\end{table}

At common uniform locations, Table~\ref{tab:measured_transfer} compares reconstruction against independent implementations of RadioUNet \cite{levie2021radiounet}, Sparse-Guided RadioUNet \cite{kojima2025sparseguided}, Saipp-Net \cite{feng2025saipp}, Physics-Feature U-Net \cite{petrosyan2025physicsfeatures}, and SIP2Net \cite{lu2025sip2net}. RadioDiff prior averages measurement-aligned candidates, and RMSampler posterior reweights them before residual correction. Full RMSampler has the lowest mean RMSE at both budgets in both datasets.

\begin{figure}[!t]
\centering
\includegraphics[width=\columnwidth]{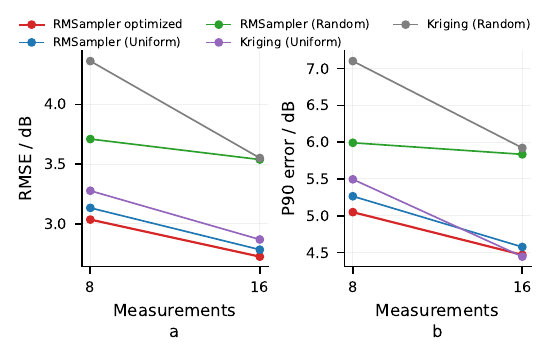}
\caption{End-to-end sampling on real-world RSRP from 10 held-out Chongqing NCI groups. Kriging uses only its own Uniform or Random locations.}
\label{fig:measured_sampling}
\end{figure}

Sampling is tested on 10 additional held-out Chongqing NCI groups with a common RMSampler reconstructor, while Kriging retains its own sampling locations. Equation~\eqref{eq:measured_acquisition} uses $q=0.3$ and $\omega_\tau=0.01$. In Fig.~\ref{fig:measured_sampling}, paired gains over random sampling are $0.67/0.81$~dB at B8/B16 with positive cell-bootstrap intervals. Gains over uniform sampling are $0.10/0.06$~dB with intervals containing zero.

\section{Conclusion}
\label{sec:conclusion}

RMSampler improves sparse RM reconstruction by coupling next-location selection with measurement-conditioned inference when the interfering emitter location is unknown. Normalized reductions in global RM and emitter-location uncertainty guide sampling, while ESS-guided correction and minimum-distortion projection use the collected measurements without changing pretrained neural parameters or architecture. Evaluations on simulated RMs and Vienna and Chongqing real-world measured data demonstrate improved reconstruction, including lower RMSE than random sampling under the common-reconstructor Chongqing protocol.

\bibliographystyle{IEEEtran}
\bibliography{rmsampler}

\end{document}